\documentclass[]{interact}

\usepackage{epstopdf}
\usepackage{subfigure}

\usepackage[numbers,sort&compress,merge]{natbib}
\bibpunct[, ]{(}{)}{,}{n}{,}{,}

\theoremstyle{plain}

\theoremstyle{definition}

\theoremstyle{remark}

\usepackage{amsmath,amsfonts,amssymb,mathrsfs,dsfont}

\begin{document}

%
%
%
%
%
%
%
%
%
%
%
%
%
%
%
%
%
%
%
%
%
%
%
%
%
%
%
%
%
%

\articletype{}

\title{Magnus expansion method for two-level atom interacting with few-cycle pulse}

\author{
\name{T.~Begzjav\textsuperscript{a}\thanks{CONTACT T.~Begzjav. Email: mn.tuguldur@physics.tamu.edu}, J. S.~Ben-Benjamin\textsuperscript{a,b}, H.~Eleuch\textsuperscript{a}, R.~Nessler\textsuperscript{a,b}, Y. Rostovtsev\textsuperscript{c}, and G.~Shchedrin\textsuperscript{d}}
\affil{\textsuperscript{a}Institute for Quantum Science and Engineering, Texas A\&M University, College Station, TX 77843, USA; \textsuperscript{b}Department of Physics, Baylor University, Waco, Texas 76798, USA; \textsuperscript{c}Department of Physics, University of North Texas, Denton, Texas 76203, USA, \textsuperscript{d}Department of Physics, Colorado School of Mines, Golden, Co. 80401, USA.}
}

\maketitle

\begin{abstract}
Using the Magnus expansion to the fourth order, we obtain analytic expressions for the atomic state of a two-level system driven by a laser pulse of arbitrary shape with small pulse area. We also determine the limitation of our obtained formulas due to limited range of convergence of the Magnus series. We compare our method to the recently developed method of Rostovtsev et al. (PRA 2009, 79, 063833) for several detunings. Our analysis shows that our technique based on the Magnus expansion can be used as a complementary method to the one in PRA 2009.
\end{abstract}


\section{Introduction}

Simple models are at the heart of fundamental physics. The harmonic oscillator in
classical mechanics, the ideal gas in statistical physics, and the two-level system in
quantum mechanics are prime examples of such models. A two-level system
(e.g.\ spin up--spin down system) driven by an electromagnetic pulse is the
quintessential problem in nuclear magnetic resonance, laser physics, and quantum
information theory \cite{Ref:Jaynes,Ref:Nielsen}.
However, simple analytical solutions are only readily available for the exactly
solvable model of a square pulse interacting with a two-level system treated
within the rotating wave approximation (RWA). In the RWA the key terms that
depend on the difference between the atomic frequency $\omega$ and the field
carrier frequency $\nu$, i.e. $\omega-\nu$, are kept while the counter-rotating terms
expressed in terms of the frequency sum $\omega+\nu$ are neglected. The usual
extension of the analytical solution for the two-level atom was to include non-RWA
terms. A number of powerful methods have been developed that treat two-level
systems beyond the RWA.

Recently, a remarkably accurate analytic solution in the case
of a two-level system interacting with a far off-resonant pulse has been found \cite{Ref:Yuri}
and applied to analyze the system's behavior due to different driving fields
\cite{Ref:Jha2, Ref:Jha}. Another way to solve the two-level problem analytically is
proposed in \cite{Ref:hichem1}. It is based on the transformation of the scattering problem into a
two-level atom, since several approximate analytical methods for the stationary
Schr\"odinger equation have shown their validity \cite{Ref:app1,Ref:app2,Ref:app3,Ref:app4,Ref:app5,Ref:app6,Ref:app7,Ref:app8,Ref:app9,Ref:app10}. However,
this approach gives practical expressions only in limited cases; in general, very
complicated expressions are generated.
Here, we obtain a new class of analytical solutions for a two-level system pumped by an arbitrarily time-dependent field of a few-cycle pulse. The present
class of solutions is based on the evolution operator technique, employing an approximation that preserves its unitarity. More precisely, we derive analytical expressions for the population dynamics of a two-level atomic system, pumped by an external field, using the Magnus expansion method. This method generates simple and surprisingly accurate solutions. The Magnus
expansion, introduced by outstanding mathematician Wilhelm Magnus in 1954
\cite{Ref:Magnus}, was applied shortly after in a variety of fields of physics, for example, for
studying nuclear spectroscopy \cite{Ref:ns1}, nuclear collisions \cite{Ref:nc1}, crystal structure \cite{Ref:cs1},
and averaging effects in magnetic resonance \cite{Ref:Haeberlen}. Nowadays, the Magnus
expansion has wide applications in several fields of physics and mathematics \cite{Ref:Mp1,Ref:Blanes,Ref:Shchedrin}.

\section{Model and calculation}
Our system of interest is a
two-level atom,
consisting of an excited state
$\vert a \rangle$
and a ground state
$\vert b \rangle$,
having an atomic transition frequency $\omega$ and interacting with an electric field.
%
%
%
%
%
%
%
%
The pulse has a frequency $\nu$ and a time-dependent Rabi frequency ${\Omega(t) = \wp E(t)/\hbar}$, where $E(t)$ represents the amplitude of the electric field and $\wp$ is the transition dipole moment.  %

In the interaction picture, the atomic state is given by%
\begin{align} %
\vert \Psi(t) \rangle = a(t) \vert a \rangle + b(t) \vert b \rangle.
\label{base1}
\end{align} %
The dynamical evolution of the wavefunction $\vert \Psi(t) \rangle $ is described by the Schr\"odinger equation %
\begin{align} %
i \hbar \frac{d}{dt} \vert \Psi(t) \rangle = H(t) \vert \Psi(t) \rangle, %
\label{base2}
\end{align} %
where the Hamiltonian $H(t)$ for the two-level system in the interaction picture has the following expression:
\begin{align} %
\label{Eq:Hamiltonian} %
H(t) = - \hbar \Omega(t)  \Big( \exp [ i \omega t]\vert a\rangle\langle b\vert + \text{h.c.} \Big). %
\end{align}
Here, without loss of generality and for simplicity,
$\Omega(t)$ is assumed to be real. %


If the initial state at $t=0$ is defined by $\vert \Psi(0)\rangle$, the formal solution at a later time $t > 0$ can be written as %
\begin{align} %
\vert \Psi(t) \rangle = U(t, 0) \vert \Psi(0) \rangle %
,
\end{align} %
where the time-evolution operator satisfies a similar equation as the state $\vert \Psi(t) \rangle$, %
\begin{align} \label{main_eqn}
i \hbar \frac{d}{dt} U(t,0) = H(t) U(t,0), %
\end{align}
and has the initial condition
$U(0,0)=1$.
To simplify notation,
we suppress the initial time $t=0$ in $U(t,0)$,
and simply write $U(t)$.

From a mathematical point of view,
Eq. \eqref{main_eqn} is a linear ordinary differential matrix equation on the complex field $\mathbb{C}$.
If the Hamiltonian $H(t)$ commutes with itself at different times ($[H(t_1), H(t_2)] = 0$),
then the time-evolution operator for Eq. \eqref{main_eqn} is
\begin{align} %
\label{Eq:TimeIndependentSolution} %
U(t) = \exp \left(-\frac{i}{\hbar} \int_0^t H(t') \, dt' \right) %
.
\end{align}
However,
the situation becomes more complicated
if the Hamiltonian does \emph{not} commute with itself
at different times.
Using standard perturbation theory,
the general solution for the time-evolution operator is

\begin{equation}\label{perturbation}
U(t)=1+\sum_{n=1}^{\infty}\left(-\frac{i}{\hbar}\right)^n\int_0^{t}dt_n\int_0^{t_n}dt_{n-1}\cdots \int_0^{t_2}dt_1 H(t_n)H(t_{n-1})\cdots H(t_1).
\end{equation}

A more compact, equivalent expression,
named after Freeman John Dyson,
is given by \cite{Ref:Dyson}
\begin{align} %
U(t) = \mathscr{T} \exp \left(-\frac{i}{\hbar} \int_0^t H(t_1) \, dt_1 \right) %
,
\end{align}
where $\mathscr{T}$ is the time-ordering operator.


%
In his seminal paper of 1954 \cite{Ref:Magnus}, Magnus claims that the general solution of the linear ordinary differential matrix equation \eqref{main_eqn} can be written as
\begin{align} %
U(t) %
= \exp \left[ \sum_{n=1}^\infty S_n(t, 0) \right] %
,
\end{align} %
and we refer to the sum in the exponent as
``the Magnus expansion''. However, as we will mention in the next section, this expansion has a limited range of validity.
The Magnus expansion method
attracts great interest among mathematicians, physicists, and chemists.
It is worth mentioning that the Magnus expansion
preserves the unitarity and symplectic property of the $U(t)$ matrix,
which is a great advantage for numerical integration methods
of linear ordinary differential equations.
The first few terms of the expansion are

\begin{subequations}
\label{magnus}
\begin{align}
S_{1}
= & \,
\frac1{(i\hbar) 1!}
\int_{0}^t dt_1 H(t_1),
\\
S_{2}
= & \,
\frac1{(i\hbar)^2 2!}
\int_{0}^t dt_1 \int_{0}^{t_1} dt_2 [ H (t_1), H(t_2)],
\\
S_{3}
= & \,
\frac1{(i\hbar)^3 3!}
\int_{0}^t dt_1 \int_{0}^{t_1} dt_2 \int_{0}^{t_2} dt_3 %
\{ [ H(t_1), [ H (t_2), H(t_3)]] + [ H(t_3), [ H (t_2), H(t_1)]] \},
\\
S_{4}
= & \,
\frac1{(i\hbar)^4 4!}
\int_{0}^t dt_1 \int_{0}^{t_1} dt_2 \int_{0}^{t_2} dt_3 \int_{0}^{t_3} dt_4 \nonumber\\ %
& \{ [[[ H(t_1), H (t_2)], H(t_3)], H(t_4)] + [ H(t_1), [[ H (t_2), H(t_3)],H(t_4)]] \nonumber \\
& +[ H(t_1), [H (t_2), [H(t_3), H(t_4)]]] + [ H(t_2), [ H (t_3), [H(t_4),H(t_1)]]] \}.
\end{align}
\end{subequations}
The explicit expression
for the operators (matrices) $S_n$
of higher order in $n$
are much more complicated,
and an explicit formula of the fifth-order Magnus expansion term
is presented in \cite{Ref:Prato} and given in the Appendix.
From an algorithmic point of view, the reference \cite{Ref:Prato} provides a formula
for finding the $n$th expansion term
from the previous terms:
\begin{align}
S_n
=
\frac1{i\hbar}
\int_0^t dt_1 \,
\left(
   H(t_1)
   -
   \frac{1}{2} [S_{n-1},H(t_1)]
   +
   \frac{1}{12} [S_{n-1},[S_{n-1},H(t_1)]]
   +
   \cdots
\right)
\end{align}

Notice that when the matrices $H(t)$ at different times commute,
the only nonzero term is $S_1$,
and the solution reduces to the well-known
Eq. \eqref{Eq:TimeIndependentSolution}. %

Motivated by \cite{Ref:Shchedrin}
we apply the Magnus expansion method to solve
the Schr\"odinger equation with the Hamiltonian given in Eq. \eqref{Eq:Hamiltonian}.
Since this Hamiltonian is off-diagonal,
and since $S_n$
involves only the summation and integration
of products of $n$ Hamiltonians at different times, %
for even $n$,
the $S_n$
are diagonal
\begin{align}
S_{2n}
=
\begin{pmatrix}
-i \phi^{(2n)} (t) & 0
\\
0 & i \phi^{(2n)} (t)
%
\end{pmatrix}
,
\end{align}
and for odd $n$,
the $S_n$ terms are
are off-diagonal
\begin{align}
S_{2n+1}
=
\begin{pmatrix}
0 & i \theta^{(2n+1)}(t)
\\
i [\theta^{(2n+1)}]^*(t) & 0
%
\end{pmatrix}
.
\end{align}
Therefore,
we write
\begin{equation}
U(t)
=
\exp{\left[\sum_{n=1}^\infty S_{n}(t)\right]}
=
\exp{\left[-i
\begin{pmatrix}
\phi(t) & -\theta(t)\\
-\theta^*(t) & -\phi(t)\\
\end{pmatrix}
\right]}.
\label{eq:01-b-11}
\end{equation}
Here,
the real-valued phase shift $\phi(t)$ is given by
\begin{equation}
\phi(t)=\phi^{(2)}(t)+\phi^{(4)}(t)+\cdots ,
\end{equation}
and the complex-valued pulse area $\theta(t)$ is
\begin{equation}
\theta(t)=\theta^{(1)}(t)+\theta^{(3)}(t)+\cdots.
\end{equation}
Note that $\phi(t)$ and $\theta(t)$ are sums of even and odd terms, since $S_n$ alternates its symmetry consecutively.
By using the formula
\begin{equation}
\exp[i({\mathbf a}\cdot{\mathbf \sigma})]=\mathds{1}\cos |{\mathbf a}| + i({\mathbf a}\cdot{\mathbf \sigma})\frac{\sin |{\mathbf a}|}{|{\mathbf a}|},
\end{equation}
where ${\mathbf \sigma}$ is the Pauli vector and $|{\mathbf a}|=\sqrt{a_1^2+a_2^2+a_3^2}$, we arrive at the final expression of the time-evolution operator
\begin{align} %
U(t)
=
\left(
\begin{array}{cc} %
\cos\beta(t) - i\displaystyle\frac{\phi(t)}{\beta(t)} \sin \beta(t)
&
i \displaystyle\frac{\theta(t)}{\beta(t)} \sin \beta(t)
\\
i \displaystyle\frac{\theta^*(t)}{\beta(t)} \sin \beta(t)
&
\cos\beta(t) + i\displaystyle\frac{\phi(t)}{\beta(t)} \sin \beta(t) %
\end{array}
\right)
,
\end{align}
where $\beta(t)$ is the real-valued magnitude
\begin{align}
\beta(t)
=
\sqrt{ \left| \theta(t) \right|^2 + \phi^2(t) }
.
\end{align}
%
Using the Hamiltonian of interest, Eq. \eqref{Eq:Hamiltonian},
and the Magnus expansion (Eqs. \eqref{magnus}),
we obtain the first two non-vanishing terms of the complex pulse area $\theta(t)$ ($\theta^{(1)}$ and $\theta^{(3)}$) as
\begin{subequations}
\begin{align} %
\theta^{(1)}(t) &= \int_0^t \Omega(t_1) \exp [ i \omega t_1] \, d t_1, %
%
\\
\mbox{and}
\nonumber
\\
%
\theta^{(3)}(t)= &
\frac{1}{3} \int_{0}^t dt_1 \int_{0}^{t_1} dt_2 \int_{0}^{t_2} dt_3\Omega(t_1)\Omega(t_2)\Omega(t_3)\nonumber\\
& \left({\rm e}^{i\omega(t_2+t_3-t_1)}+{\rm e}^{i\omega(t_1+t_2-t_3)}-2{\rm e}^{i\omega(t_1+t_3-t_2)}\right)
,
\end{align}
\end{subequations}
while the first two non-vanishing contributions to the phase shift ($\phi^{(2)}$ and $\phi^{(4)}$) can be written as
\begin{subequations}
\begin{align} %
&\phi^{(2)}(t) = \int_0^t dt_1 \int_0^{t_1} d t_2 \, \Omega(t_1) \Omega(t_2) \sin [ \omega ( t_1 - t_2) ], %
%
\\
\mbox{and}
\nonumber
\\
%
&\phi^{(4)}(t)
=
-\frac{2}{3}\int_{0}^t dt_1 \int_{0}^{t_1} dt_2 \int_{0}^{t_2} dt_3 \int_{0}^{t_3} dt_4 \nonumber\\
&\Omega(t_1)\Omega(t_2)\Omega(t_3)\Omega(t_4)\cos(\omega (t_4-t_1)\sin(\omega(t_3-t_2)))
.
\end{align}
\end{subequations}
%

%
%
%
%
%

%
In order to develop an analytical approximation for the two-level system, we truncate the Magnus expansion $\sum S_n$
to both second order and fourth order.
We insert the truncated Magnus expansions into Eq. \eqref{eq:01-b-11}
to find two approximations for the time-evolution operator,
$U^{(2)}(t)$ and $U^{(4)}(t)$,
respectively.
We use our approximate time-evolution operators
to evolve the state $\left| \Psi(0) \right\rangle$,
and we will compare in the next section the results with
those obtained by fourth-order perturbation theory,
and with those obtained by numerics.

Using our $U^{(2)}(t)$, and placing the atomic wavefunction initially in the ground state ${\left| \Psi(0) \right\rangle = \left| b \right\rangle}$, we obtain

\begin{align}\label{2nd} %
\vert \Psi^{(2)}(t) \rangle %
&\approx \left[ i \frac{\theta^{(1)}(t)}{\beta^{(2)}(t)} \sin \beta^{(2)}(t) \right] \vert a \rangle\nonumber\\
&+ \left[ \cos \beta^{(2)}(t) + i \frac{\theta^{(1)}(t)}{\beta^{(2)}(t)} \sin \beta^{(2)}(t) \right] \vert b \rangle.
\end{align}
Applying instead our $U^{(4)}(t)$ we find that
\begin{align}\label{4th} %
\vert \Psi^{(4)}(t) \rangle %
&\approx \left[ i \frac{\theta^{(1)}(t)+\theta^{(3)}(t)}{\beta^{(4)}(t)} \sin \beta^{(4)}(t) \right] \vert a \rangle\nonumber\\ + &\left[ \cos \beta^{(4)}(t) + i \frac{\theta^{(1)}(t)+\theta^{(3)}(t)}{\beta^{(4)}(t)} \sin \beta^{(4)}(t) \right] \vert b \rangle. %
\end{align}
Here,
the $\beta$'s are
\begin{align} %
\beta^{(2)}(t) = \sqrt{ \vert \theta^{(1)}(t) \vert^2 + (\phi^{(2)}(t))^2} %
\end{align}
and
\begin{align} %
\beta^{(4)}(t) =\sqrt{ \vert \theta^{(1)}(t)+\theta^{(3)}(t) \vert^2 + (\phi^{(2)}(t)+\phi^{(4)}(t))^2}. %
\end{align}
These are the approximate solutions of the two-level atom interacting with a laser pulse of an arbitrary shape.
Before proceeding to our numerical analysis,
we discuss here
the convergence, and implications of our results.
In Magnus's original paper,
the issue of convergence is not considered.
But it has attracted great attention and has been extensively studied for the past half-century.
In general,
the Magnus expansion converges
only in a limited time interval.
The interval of convergence, $0\leq t\leq T$,
depends on
the Frobenius norm \cite{Ref:Golub} of the Hamiltonian $H(t)$,
and can be deduced from the inequality \cite{Ref:Blanes}
\begin{align} %
\label{Eq:ConvergentRange} %
\int_0^T \lVert -\frac{i}{\hbar}H(t)\rVert\,dt < r_c
,
\end{align}
where $\lVert\cdot\rVert$ stands for the Frobenius norm and $r_c$ is a real number. %
%


To find the convergence criterion for our situation,
we use the Hamiltonian Eq. \eqref{Eq:Hamiltonian}
and Eq. \eqref{Eq:ConvergentRange}
and find that the inequality
\begin{align} %
\label{Eq:ConvergentRangeSimplified} %
\int_0^T\vert \Omega(t)\vert\, dt < \frac{r_c}{\sqrt{2}},
\end{align}
 must be satisfied. This raises an obvious question: how is $r_c$ calculated?
Several values for $r_c$ are found in the literature.
For example, Pechukas and Light \cite{Ref:Pechukas} have found that $r_c=\log 2$, while
S.\ Blanes et al.\ \cite{Ref:Blanes2} calculate $r_c=1.08686$.
Later, Moan and Niesen \cite{Ref:Moan} provide $r_c=\pi$,
and show that the restriction in Eq.\ \eqref{Eq:ConvergentRange} is not strict;
in other words,
it gives only an approximate value for convergence domain.
From a physical point of view,
if we use the value $r_c=\pi$,
the restriction in Eq. \eqref{Eq:ConvergentRangeSimplified} means that the solutions in
Eqs.\eqref{2nd} and \eqref{4th} are valid
for weak pulse areas of roughly less than $\pi/\sqrt{2}$,
though it is unclear whether there is a strict limit on the pulse area.

\section{Numerical analysis}
In this section we apply our approximate solutions, Eqs.\ \eqref{2nd} and \eqref{4th}, to a Gaussian pulse driving the two-level system described as
\begin{align} %
\Omega(t) = \Omega_0\exp \big(-a ( t - \tau)^2 \big) \cos ( \nu (t-\tau)), %
\label{eq:b-04-29}
\end{align}
where $\nu$ represents the frequency of the pulse and
$\Omega_0$ is its amplitude.
In order to test the convergence of the Magnus expansion
and its dependence on the pulse area,
we consider three pulses of different areas:
one weak pulse of an area $A=\pi/20$,
according to
\begin{align}
A
=
\int_{-\infty}^\infty dt \,\lvert\Omega(t)\rvert
\end{align}
which is less than the boundary value $\pi/\sqrt{2}$
for the Magnus method,
and pulses of area $\pi/2$ and $\pi/\sqrt{2}$,
and compute the time evolution of the two-level system for each pulse.

\begin{figure}[h!] %
\begin{center}
\includegraphics[scale=1.4]{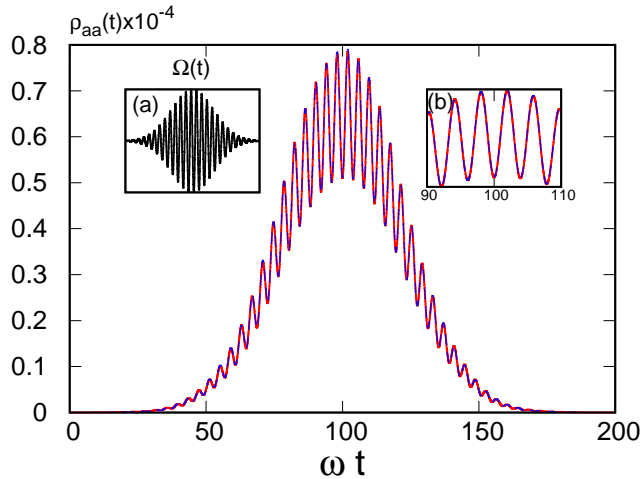}
\caption{
The excited-state population is plotted as a function of time
for a weak $\pi/20$ Gaussian envelope pulse of frequency $\nu=0.8\omega$.
The results of our numerical simulation are plotted as a solid red line,
and the results of
our 4th-order Magnus expansion result (Eq. \eqref{4th})
are overlaid as dashed blue line.
In inset (a), we show the pulse profile in the time domain,
and in inset (b), we magnify the main plot (of the excited-state population)
in the interval $90\leq t\leq 110$.
The parameters we use are
$\Omega_0=0.0038937\omega$,
$a=0.0005\omega^2$, and
$\tau=100\omega^{-1}$.
}
\label{Fig:weak}
\end{center}
\end{figure}

\subsection{Weak pulse}
First,
for a weak Gaussian pulse with pulse area $\pi/20$,
the excited-state population is calculated using both
4th order Magnus expansion method and standard 4th order Runge--Kutta integration method.
The results are shown in Figure \ref{Fig:weak}.
Frequency and time units are in atomic transition frequency $\omega$ and its inverse $\omega^{-1}$ respectively.
We choose parameters which are from usual experimental situations.
We take the atomic frequency $\omega=10^{15}\text{ s}^{-1}$, and the Rabi frequency of the pulse is calculated to be $\Omega_0=0.0038937\omega\sim 10^{12}\text{ s}^{-1}$.
We take a parameter $a=0.0005\omega^2$,
which corresponds to $\text{FWHM}\approx 80\text{fs}$.
As we mentioned in a previous section, our Magnus expansion method converges well
in the case of a weak pulse.

\subsection{Strong short pulse}
\begin{figure}[h!]
\begin{center}
\includegraphics[scale=1.0]{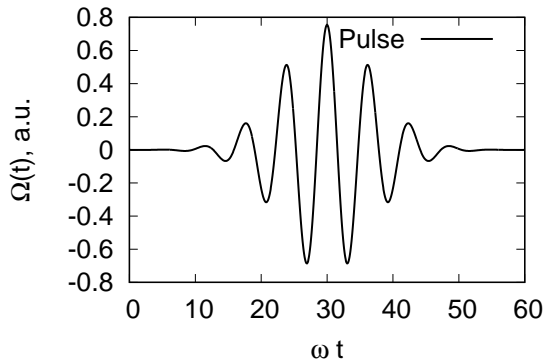}
\caption{Shape of the pulse used in the numerical calculation of Fig. 3 and 4 .}\label{Fig:z1}
\end{center}
\end{figure}
\begin{figure}[h!]
\begin{center}
\includegraphics[scale=1.0]{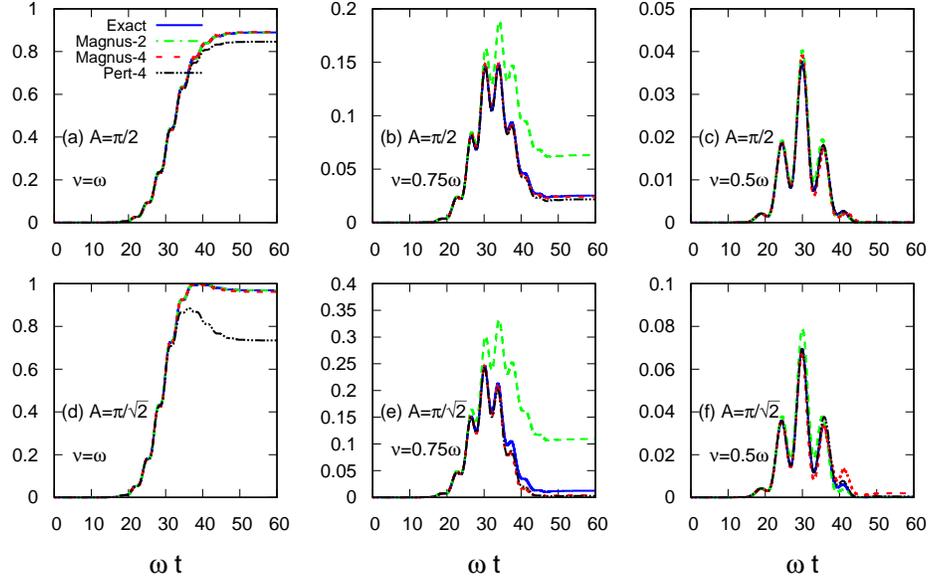}
\caption{Atomic excited state population as a function of $\omega t$. The pulse area and detuning are indicated in each plot. Other parameters are $a=0.01\omega^2$ and $\tau=30\omega^{-1}$. The legend for the colors and line types, shown in plot (a), applies to all plots.}\label{Fig:z2}
\end{center}
\end{figure}

Next we study the possibility of applying the Magnus expansion method in the strong pulse regime. Using three different methods--our Magnus expansion methods of second and fourth orders, the perturbation methods of fourth order and the fourth order Runge-Kutta numerical integration---we calculated the time evolution of the two-level atom driven by a few-cycle pulse of the form of Eq.\eqref{eq:b-04-29} with $a=0.01\omega^2$ ($FWHM\approx 16.6\text{fs}$ for $\omega=10^{15}s^{-1}$) and different detunings $\Delta=\omega-\nu$ for the pulse area $\pi/2$ and $\pi/\sqrt{2}$. It is worth mentioning that few ($\sim 6$) cycles are contained in our laser pulse (see Figure \ref{Fig:z1}). The dynamics of the excited-state populations determined by these methods are plotted in the Fig. \ref{Fig:z2}.

As shown in Figure \ref{Fig:z2},
the time evolution of the two-level system
driven by the few-cycle pulse of area less than $\pi/\sqrt{2}$ is well described by the fourth order Magnus expansion method but not by the second order Magnus expansion.
 Because perturbation theory does not conserve the unitarity for low order, it cannot describe any strong atom--field interaction (see plots (a) and (d)).
 When the pulse area increases beyond $\pi/\sqrt{2}$ the validity of the Magnus fourth order method is not guaranteed.

\begin{figure}[h!]
\begin{center}
\includegraphics[scale=1.0]{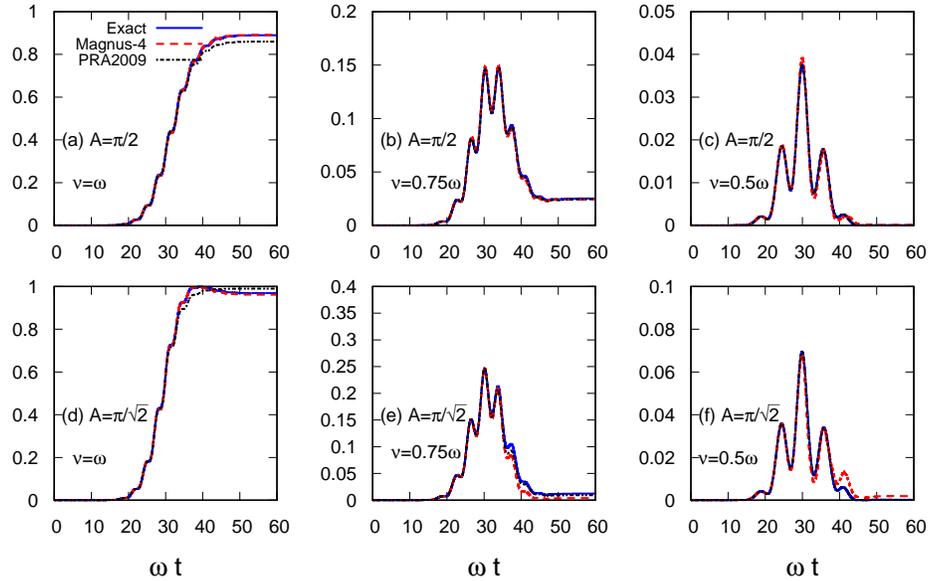}
\caption{Atomic excited state population as a function of $\omega t$ calculated using method in the paper \cite{Ref:Yuri} is compared to Magnus 4th order and 4th order Runge--Kutta method. The pulse area and detuning are shown in each plot. Other parameters are $a=0.01\omega^2$ and $\tau=30\omega^{-1}$. The legend for the colors and line types, shown in plot (a), applies to all plots.}\label{Fig:z3}
\end{center}
\end{figure}

To compare our method to the method developed in the paper \cite{Ref:Yuri}, we have plotted the time evolution of the excited-state population in Figure \ref{Fig:z3} using the same areas and detunings as Figure \ref{Fig:z2}. Plots (b), (c), (e), and (f) demonstrate that method \cite{Ref:Yuri} works very well for large area or when the population in the excited state is smaller than that in the ground state. Meanwhile, 4th order Magnus expansion method works very well in the case of small area and when the atom-field interaction is strong, meaning that the excited state is highly populated during the interaction.
 The plots (a) and (f) in Figure \ref{Fig:z3} indicate that these two methods, namely, 4th order Magnus expansion and the method in the paper \cite{Ref:Yuri}, are complementary.

\section{Conclusions}

We have derived analytical solutions based on the Magnus expansion for the time evolution of a two-level system excited by an external time-dependent electric field. Our method goes beyond the rotating wave approximation and applies to a two-level atom interacting with an arbitrary-shaped laser pulse. We have also shown that our method performs better than other methods for an ultrashort pulse. Our approximate expressions work well for a pulse area below $\pi/\sqrt{2}$ for any detuning, but it is unclear whether this restriction, due to the finite convergence interval of the Magnus expansion, is strict, since more precise convergence criteria have not yet been found. We have also observed that the method developed in \cite{Ref:Yuri} works very well for large area pulse. In the sense of their applicable parameter range, we can consider that these two methods are complementary analytical techniques for describing the dynamics of the two-level system excited by a variable pulse.


\section*{Acknowledgement(s)}
The authors would like to thank Moochan Kim, Petr Anisimov, David Lee, Marlan Scully and Wolfgang Schleich for helpful discussions. T.\ B. is supported by the Herman F.\ Heep and Minnie Belle Heep Texas A\&M University Endowed Fund held/administered by the Texas A\&M Foundation.

\section*{Funding}
Office of Naval Research (Award No. N00014-16-1-3054); Robert A. Welch Foundation (Grant No. A-1261).

\setcounter{secnumdepth}{0}
\section{Appendix. Explicit expression for the 5th order Magnus term}\label{app:a}
Fifth order Magnus expansion term \citep{Ref:Prato} is explicitly
\begin{align*}
S_5
=&
\frac{2i}{(i\hbar)^5 5!}
\int_0^t dt_1\int_0^{t_1} dt_2\int_0^{t_2} dt_3\int_0^{t_3} dt_4\int_0^{t_4} dt_5
\nonumber
\\
& \left\{ -2[H(t_5),[H(t_4),[H(t_3),[H(t_2),H(t_1)]]]]+8[H(t_1),[H(t_5),[H(t_4),[H(t_2),H(t_3)]]]] \right.\nonumber\\
& \left. +4[[H(t_5),H(t_1)],[H(t_4),[H(t_2),H(t_3)]]]+4[[H(t_4),H(t_1)],[H(t_5),[H(t_2),H(t_3)]]]\right.\nonumber\\
& \left. -[[H(t_2),H(t_3)],[H(t_5),[H(t_4),H(t_1)]]]+4[[H(t_3),H(t_1)],[H(t_5),[H(t_2),H(t_4)]]]\right.\nonumber\\
& \left. -[[H(t_2),H(t_4)],[H(t_5),[H(t_3),H(t_1)]]]-[[H(t_2),H(t_5)],[H(t_4),[H(t_3),H(t_1)]]]\right.\nonumber\\
& \left. -[[H(t_3),H(t_4)],[H(t_5),[H(t_2),H(t_1)]]]-[[H(t_3),H(t_4)],[H(t_1),[H(t_2),H(t_5)]]]\right.\nonumber\\
& \left. -[[H(t_5),H(t_1)],[H(t_3),[H(t_2),H(t_4)]]]-[[H(t_4),H(t_1)],[H(t_3),[H(t_2),H(t_5)]]]\right.\nonumber\\
& \left. -[[H(t_3),H(t_5)],[H(t_4),[H(t_2),H(t_1)]]]-[[H(t_3),H(t_5)],[H(t_1),[H(t_2),H(t_4)]]]\right.\nonumber\\
& \left. -2[[H(t_1),[H(t_4),[H(t_3),[H(t_2),H(t_5)]]]]-[[H(t_4),H(t_5)],[H(t_1),[H(t_2),H(t_3)]]]\right.\nonumber\\
& \left. -[[H(t_2),H(t_3)],[H(t_1),[H(t_4),H(t_5)]]]-[[H(t_2),H(t_4)],[H(t_1),[H(t_3),H(t_5)]]]\right.\nonumber\\
& \left. -[[H(t_2),H(t_1)],[H(t_4),[H(t_3),H(t_5)]]]-[[H(t_4),H(t_5)],[H(t_3),[H(t_2),H(t_1)]]]\right.\nonumber\\
& \left. -[[H(t_3),H(t_1)],[H(t_4),[H(t_2),H(t_5)]]]-2[H(t_1),[H(t_5),[H(t_3),[H(t_2),H(t_4)]]]]\right\}\nonumber\\
\end{align*}
%
We note that
each term in the integral involves four commutators.

\bibliographystyle{tfp}
\bibliography{Manuscript}

\end{document}